\documentclass{osa-article}

\journal{oe}

\usepackage{amsfonts}
\usepackage{amsmath}
\usepackage{amssymb}
\usepackage{graphicx}
\usepackage{hyperref}

\begin{document}

\title{Noise-induced temporal regularity and signal amplification in an optomechanical system with parametric instability}
\author{Danying Yu, Min Xie, Yanbei Cheng, and Bixuan Fan\authormark{*}}
\address{College of Physics and Communication Electronics, Jiangxi Normal University, Nanchang, 330022, China}

\email{\authormark{*}fanbixuan@jxnu.edu.cn}

\begin{abstract}
Noise usually has an unwelcome influence on system performance. For instance, noise inevitably affects the low-frequency mechanical freedom in optomechanical experiments. However, we investigate here the beneficial effects of thermal noise on a basic optomechanical system with parametric instability. In a regime near parametric instability, it is found that thermal noise in the mechanical freedom can sustain long-term quasi-coherent oscillations when the system would otherwise remain in the equilibrium state. In an overlapping regime of parametric instability and bistability, intermittent switching between a self-sustained oscillating state and an equilibrium can be induced by adding a certain amount of noise. When a subthreshold periodic signal is applied to the mechanics, the switching between the self-sustained oscillations and the equilibrium exhibits good periodicity at a rate that is synchronized to the signal frequency, resulting in a significant amplification of the signal. Our results deepen the understanding of the interplay between optomechanical nonlinearity and noise and provide theoretical guidance for experimental observation of noise-induced beneficial phenomena in optomechanics.
\end{abstract}

\section{Introduction}

The field of cavity optomechanics has developed rapidly in the past few decades. As a solid-state quantum platform, the optomechanical system has exhibited both great potential for testing fundamental problems in quantum mechanics and wide applications in quantum information processing and quantum precise measurement (see reviews \cite{OM_rev0,OM_rev,OM_rev2}). Because of the radiation-pressure-induced nonlinearity between optics and mechanics, cavity optomechanical systems allow the study of a variety of nonlinear phenomena, such as bistability \cite{Samuel,Bist}, multi-stability \cite{LC_OM1,MultiS1,Fan}, self-sustained oscillations \cite{LC_OM1,Zhang2014,LC_OM5}, and chaotic motion \cite{Chaos1,Chaos2}. However, noise is unavoidable in open optomechanical systems, and compared to all-optical systems the mechanical freedom in an optomechanical system is more sensitive to thermal noise because of its relatively low resonance frequency. Thus, the effect of noise on optomechanical nonlinearity has become an important topic \cite{Aldana,Suchoi}.

In nonlinear systems, noise and random fluctuations do not always act destructively. Some physical mechanisms allow noise to act beneficially, such as stochastic resonance (SR) \cite{Benzi81,Benzi82} and coherence resonance (CR) \cite{CR1,CR0,CR2}. As a mechanism for enhancing the response of a nonlinear system to a weak external signal, SR has been observed and investigated in various fields (see review \cite{Gammaitoni}) and used widely to detect \cite{Sun2015,Wang2017,Shao2018},  amplify \cite{amp1,amp2}, and reconstruct \cite{Cao14,Han17} weak signals. Recently, SR has been introduced in the field of cavity optomechanics, examples being chaos-mediated optomechanical SR \cite{Monifi} and SR in a tristable optomechanical system \cite{Fan}. However, although optomechanical SR is now being investigated, there are more aspects to be explored. CR is the phenomenon of purely noise-induced temporal regularity, which has been observed in excitable \cite{CR0,Lee2010} and nonexcitable \cite{Liu2001,CR2} systems but is yet to be reported in the context of an optomechanical system.

In the present work, we study the constructive effects of thermal noise on the nonlinear behavior of a basic optomechanical system with self-sustained oscillations. Our investigation comprises two parts, the first being a study of purely noise-induced novel dynamics in the form of (i) noise-sustained quasi-coherent oscillations near the Hopf bifurcation and (ii) noise-induced random switching between self-sustained oscillations and equilibrium in the overlapping regime of self-sustained oscillations and bistability. In particular, (i) is often referred to as CR, and the present investigation may be the first of optomechanical CR. In the second part, we study how a subthreshold periodic signal and thermal noise affect system performance. We find that large-amplitude periodic switching between self-sustained oscillations and equilibrium can be observed numerically, indicating the occurrence of SR and an amplification of the external signal by noise. More interestingly, there is a double-peak structure in the power spectrum, which differs from typical SR behavior. Based on experimentally feasible conditions, our results will help to realize optomechanical SR and CR experimentally in a single setup.

The rest of this paper is structured as follows. In the Sec. 2, we introduce the model and give linear stability analysis of our system. In Sec. 3, we study noise induced system dynamics in the absence of an external signal, including the CR phenomenon and noise induced state switching. In Sec. 4, we study noise induced system dynamics in the presence of a weak periodic force, in particular, the stochastic resonance phenomenon with self-sustained oscillations. Finally, we conclude our paper in Sec. 5.

\begin{figure*}
  \centering
  \includegraphics[width=5in]{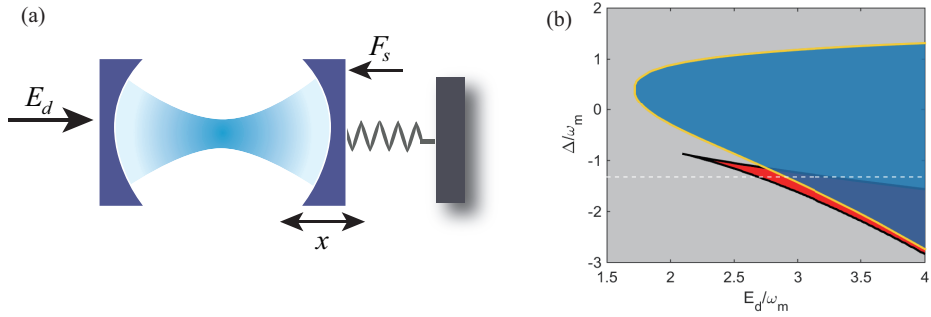}
  \caption{(a) Optomechanical system driven by a coherent optical field $E_d$ and a periodic force signal $F_s$. (b) System stability diagram. The parameter plane is divided into four regions: (i) one stable fixed point (grey); (ii) three fixed points (red); (iii) parametric instability (blue); (iv) overlap region (purple).  The white dashed line indicates the detuning ($\Delta=-1.38\omega_m$ ) used in the following numerical calculations. The system parameters are $(g,\kappa,\gamma_m) =  (0.21,1,0.25) \times \omega_m.$}
\label{fig1}
\end{figure*}

\section{Model and theory}
We consider a basic optomechanical system, namely a driven optical cavity with a freely oscillating end-mirror; see Fig.~\ref{fig1}(a). The optical mode inside the cavity is coupled with the mechanical oscillation mode of the mirror via radiation pressure. The cavity is driven coherently by a field with amplitude $E_d$ and frequency $\omega_d$. A weak force $F_s$ with frequency $\omega_f$ acts on the oscillating mirror as a probe signal. In the weak driving regime, i.e., the radiation pressure induced mirror displacement is small compared to the cavity length, nonlinear modulation on the cavity frequency by the mirror displacement can be ignored \cite{law95,Zaitsev2012,OM_rev}. Thus, the system Hamiltonian after rotating at the driving frequency $\omega_d$ can be approximately given by (with $\hbar=1$)
\begin{eqnarray}
\label{H}\nonumber
\hat{H} &=&\frac{1}{2}\omega _{m}\left( \hat{x}^{2}+\hat{p}^{2}\right) -\Delta\hat{a}^\dag\hat{a}-g\hat{a}^{\dagger }\hat{a}\hat{x}+iE_{d}(\hat{a}-\hat{a}^\dag)\\
&+&F_s \hat{x}\cos(\omega_f t),
\end{eqnarray}%
where $\hat{a}^{\dagger }\left( \hat{a}\right) $ is the creation (annihilation) operator of the optical mode; $\hat{x}$ and $\hat{p}$ are the dimensionless position and momentum operators, respectively, of the mechanical mode, which satisfy $[\hat{x},\hat{p}]=i$; $\omega_m$ is the mechanical resonance frequency; $g$ is the optomechanical coupling coefficient; and $\Delta=\omega_c-\omega_d$ is the detuning between the driving field and the cavity resonance frequency $\omega_c$. The equations of motion for the system operators are then
\begin{eqnarray}
\frac{d\hat{a}}{dt} &=&-( -i\Delta +\kappa/2  -ig\hat{x})\hat{a}-E_{d}, \\
\frac{d\hat{a}^\dag}{dt} &=&-( i\Delta +\kappa/2  +ig\hat{x})\hat{a}^\dag-E_{d}, \\
\frac{d\hat{p}}{dt} &=&-\gamma_m\hat{p}-\omega _{m}\hat{x}+g\hat{a}^\dag\hat{a}-F_s\cos(\omega_f t)+\xi,\\
\frac{d\hat{x}}{dt} &=&\omega _{m}\hat{p}.
\end{eqnarray}
Here, the optical decay rate $\kappa$ and the mechanical decay rate $\gamma_m$ are introduced phenomenologically. $\xi$ is the thermal white noise in the mechanical mode and satisfies
$\left\langle {\hat \xi (t){{\hat \xi }^ \dag }(t')} \right\rangle=2D_m \delta (t - t')$ and
$\left\langle {\xi (t)} \right\rangle=0$. Note that any effect of thermal noise on the optical mode is neglected because $k_B T\ll \hbar \omega_c$ for room temperature and below.

We begin by studying the stability properties of the system in the absence of the weak signal $F_s$. By setting the time derivatives in the equations of motion to zero, we easily obtain the steady-state equation for the mechanical freedom, namely
\begin{eqnarray}
\omega_m(\kappa^2/4+(\Delta+gx_s)^2)x_s-gE^2_d=0.
\end{eqnarray}
This is a cubic equation of the steady-state mechanical position $x_s$ and has three real roots at most. The stability of these solutions derived above can be investigated by evaluating the eigenvalues of the associated Jacobian matrix, i.e., the solution is stable if the real part of each eigenvalue is negative. Therefore, we linearize the system operators around their steady-state values, namely $\hat{a} \rightarrow \alpha_s+\delta\hat{a}$, $\hat{a}^\dag\rightarrow \alpha^*_s +\delta\hat{a}^\dag$, $ \hat{x}\rightarrow x_s +\delta\hat{x}$, and $\hat{p}\rightarrow p_s +\delta\hat{p}$. By ignoring higher-order terms, we have the linearized equation of motion as
\begin{eqnarray}
\dot{y} = J\hat{y}+\xi_0,
\end{eqnarray}
where $\hat{y}=[\delta\hat{a},\delta\hat{a}^\dag,\delta\hat{p},\delta\hat{x}]^T$, $\xi_0=[0, 0,\xi,0]^T$, and  the Jacobian matrix $J$ is given by
\begin{equation}
J=
\left (\begin{matrix}
i(\Delta+gx_s)-\frac{\kappa}{2}{\rm{ }}&0&0& {\rm{ }}ig\alpha_s\\
0&{\rm{ }}-i{\rm{(}}\Delta{\rm{ + }}gx_s{\rm{)}}-\frac{\kappa}{2}&0& -ig{\alpha^*_s}\\
g{\alpha^*_s}&g\alpha_s&-\gamma_m &-\omega _m\\
0&0&\omega_m &0
\end{matrix}\right).
\end{equation}
The eigenvalues of $J$ are the roots of the characteristic polynomial $\det(\lambda I_4-J)={\lambda ^4} + {c_1}{\lambda ^3} + {c_2}{\lambda ^2} + {c_3}\lambda  + {c_4}$,
where
\begin{eqnarray}
{c_1} &=&\gamma_m  + \kappa,  \\
{c_2} &=& {(gx_s + \Delta )^2} + \frac{1}{4}{\kappa ^2} + \gamma_m \kappa  + {\omega ^2_m}, \\
{c_3} &=& \gamma_m {(gx_s + \Delta )^2} + \frac{1}{4}\gamma_m {\kappa ^2} + \kappa {\omega ^2_m}, \\
{c_4} &=& ({(gx_s + \Delta )^2} + 2gx_s(\Delta  + x_sg) + \frac{1}{4}{\kappa ^2}){\omega ^2_m}.
\end{eqnarray}

It is usually difficult to compute the eigenvalues of $J$ directly. An easier method is to determine the stability properties by evaluating the coefficients $c_1$ to $c_4$ using the Routh--Hurwitz stability criterion \cite{RHC}. Thus, we write the Hurwitz determinants as
\begin{eqnarray}
D_1&=&
 c_1,\\
D _2&=&
\left|\begin{matrix}
c_1 & c_3\\
1 & c_2
\end{matrix}\right|,\\
D _3&=&
\left|\begin{matrix}
c_1 & c_3 & 0\\
1 & c_2 & c_4\\
0 & c_1 & c_3
\end{matrix}\right|,\\
D _4&=&
\left|\begin{matrix}
c_1 & c_3 & 0 & 0\\
1 & c_2 & c_4 &0\\
0 & c_1 & c_3 &0\\
0 & 1 & c_2 & c_4
\end{matrix}\right|=c_4 D_3.
\end{eqnarray}

According to the Routh--Hurwitz stability criterion, our system is stable if and only if all Hurwitz determinants are positive, i.e., $D_{j} (j=1,2,3,4)>0$. Furthermore, determine the existence of self-sustained oscillations, we employ  Liu's criterion \cite{Liu} for a simple Hopf bifurcation (i.e., a transition from a fixed point to a limit cycle), whereby two roots of the characteristic polynomial cross the imaginary axis and all other roots have negative real parts. Applying Liu's Hopf-bifurcation criterion to our model, a Hopf bifurcation  occurs when the following conditions are satisfied:
\begin{eqnarray}
D_3 =0, D_1>0, D_2>0, c_4>0, \rm{and} \: \frac{d D_3}{d\lambda}>0.
\end{eqnarray}
Based on these criteria, we obtain the stability diagram of our system in the $E_d$--$\Delta$ plane as shown in Fig.~\ref{fig1}(b). The plane is divided into several regions with different stability properties. The grey region contains one stable fixed point. The red region inside the black curve and outside the orange curve is bistable, containing two stable fixed points and one unstable fixed point. The blue region inside the orange curve is a region of parametric instability in which the system exhibits limit-cycle, period-doubling, and chaotic behaviors as the driving strength is increased. The most complicated region is the overlap between bistability and parametric instability, in which one stable fixed point (the lowest one), one unstable fixed point (the middle one), and one limit cycle or chaotic solution (the highest one) coexist. This overlapping region and its vicinity comprise the parameter regime of interest herein.

\section{System response to pure noise}
In this section, we study purely noise-induced novel dynamics in our system near and inside the overlapping regime of bistability and parametric instability.

\begin{figure*}
  \centering
  \includegraphics[width=5in]{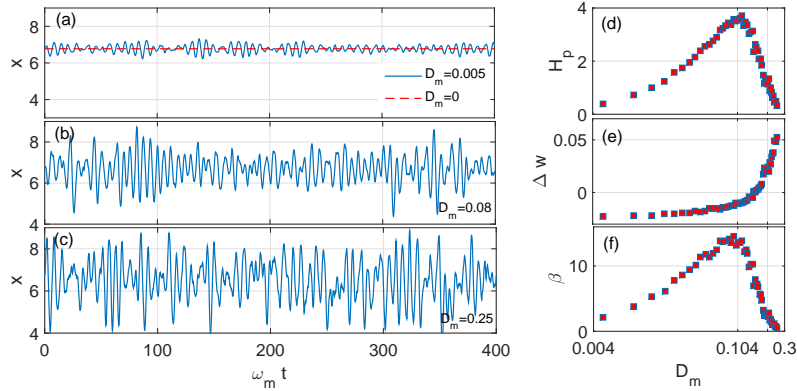}
  \caption{Noise-sustained quasi-coherent oscillations. (a--c) Single trajectories of system dynamics (the mechanical position $x$) with different noise strengths. (d--f) Peak height $H_\omega$, peak width $\Delta_\omega$, and coherence factor $\beta$ as functions of noise strength $D_m$. The system parameters are $(g,\kappa,\gamma,E_d,\Delta) =  (0.21, 1,0.25, 2.85,-1.38) \times \omega_m$.}\label{CR}
\end{figure*}

We begin by considering the situation in which the driving strength $E_d$ is below the Hopf bifurcation, meaning that there is no self-sustained oscillation. According to the bifurcation diagram in Fig.~1(b), when the detuning is set at $\Delta=-1.38\omega_m$ [labelled by the white dashed line in Fig.~1(b)] the bifurcation occurs at about $E_d=3\omega_m$, below which the system has one or three fixed points and above which the system experiences self-sustained oscillation. First, we take the driving strength to be $E_d=2.85\omega_m$. Under these conditions and in the absence of noise, a system variable (specifically the mechanical position $x$) already at its equilibrium value will remain so [see Fig.~\ref{CR}(a)]. When relatively low noise ($D_m=0.005$) is introduced in the system, the mechanical position undergoes a quasi-coherent small-amplitude oscillation with certain variations in the oscillation amplitude. As the noise amplitude $D_m$ is increased to an intermediate value (i.e., $D_m=0.08$), the amplitude of the noise-sustained oscillation grows and the regularity decreases slightly. If we continue to increase the noise strength, the oscillation amplitude increases further and the regularity worsens further. This type of noise-induced quasi-coherent oscillation is usually referred to as CR and has been observed in a wide range of systems including excitable and non-excitable ones. However, to the best of our knowledge it has not been reported in an optomechanical system until now, making the present study the first example of CR in an optomechanical system.

To better understand the CR in our model, we evaluate the peak width ($\Delta_\omega$), the peak height ($H_\omega$), and the coherence factor $\beta=\omega H_\omega/\Delta_\omega$ \cite{CR2,PhysRevE.92.042909} from the spectra of the mechanical position while varying the noise strength. As shown in Fig.~\ref{CR}, the spectral peak widens monotonically as the noise increases, while there is a clear mono-peak structure in the curves of height $H_\omega$ and coherence factor $\beta$, indicating the existence of resonance behavior that depends on the noise strength. The optimal noise corresponding to the best coherence factor is $D_m \approx 0.08$, which is consistent with the result in the time domain [Fig.~\ref{CR}(a)].

\begin{figure*}
  \centering
  \includegraphics[width=5in]{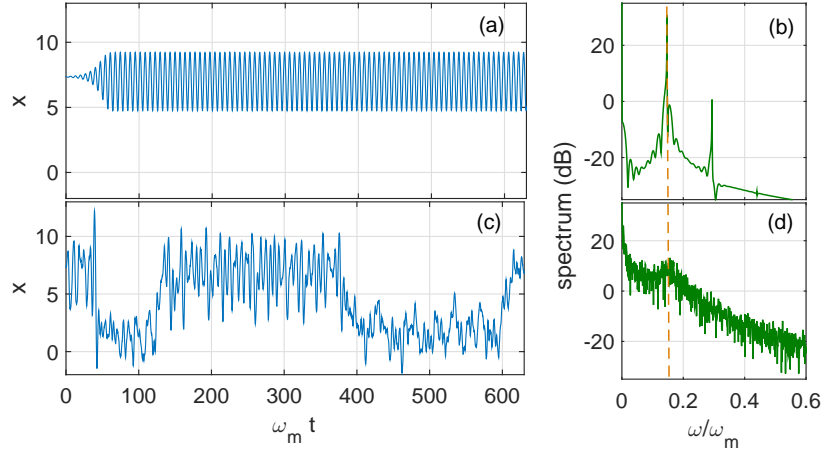}
  \caption{ Time series of system variable ($x$) and their power spectra with (a,b) no thermal noise ($D_m=0$) and (c,d) thermal noise ($D_m=0.55$). The system parameters are the same as in Fig.~\ref{CR} except for $E_d=3.11\omega_m$.}\label{switch}
\end{figure*}

We turn next to how noise affects the system dynamics inside the overlapping regime described above. We assume that initially the mechanical position is on the upper branch of the bistable structure. Here, ``bistability'' does not mean that there are two branches with stable fixed points. Instead, the upper branch here corresponds to limit-cycle dynamics or chaotic motion. Without noise, the mechanical position $x$ oscillates coherently [Fig.~\ref{switch}(a)]. In its spectrum [Fig.~\ref{switch}(b)], apart from the main peak associated with the limit cycle, there are peaks at double and higher frequencies that arise from period doubling. After adding a certain level of mechanical noise ($D_m=0.5$) to the system, random switching between the upper and lower branches is observed numerically, that is similar to other noise-induced bistable transitions. However, upon zooming in, we find that the dynamics of the high-amplitude state have a certain periodicity, which is verified by the single-peak power spectrum in Fig.~4(b). Interestingly, the inclusion of noise suppresses the double- and higher-frequency components as evidenced by only a single peak remaining in the spectrum, corresponding to the main peak in the noise-free case. These phenomena are evidence for a stochastic transition between the self-sustained oscillations and the fixed-point dynamics. Similar noise-induced intermittency between limit-cycle and fixed-point dynamics has been observed experimentally in optomechanics \cite{Suchoi}. This transition provides a basis for the occurrence of SR between self-sustained oscillations and fixed-point dynamics as discussed in the next section.

\section{System response to noise and an external signal}

\begin{figure*}
  \centering
  \includegraphics[width=5in]{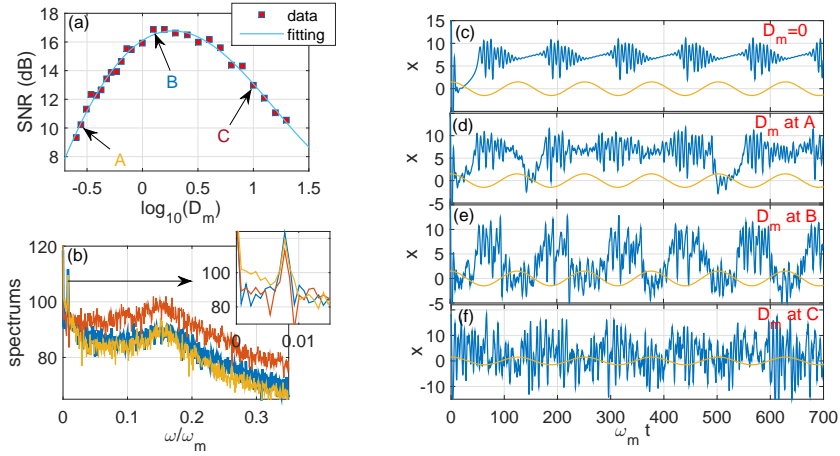}
  \caption{Stochastic resonance (SR) with self-sustained oscillations. (a) Signal-to-noise ratio (SNR) versus noise strength $D_m$. The red squares are the original data for the SNR calculated from trajectories and the blue curve is a fit to the SNR data. Points A, B, and C are representative of the low-noise, optimal, and high-noise regimes, respectively. (b) Spectra at noise levels A (yellow), B (blue), and C (red). (c) Time series of mechanical position $x$ with no noise. (d--f) Typical trajectories of mechanical position $x$ at noise levels A--C, respectively. The parameters are $(g,\kappa,\gamma,E_d,\Delta,F_s,f_s) =  (0.21, 1,0.25, 3.11,-1.38,1.5,0.05) \times \omega_m$.}
  \label{SR}
\end{figure*}

In this section, we study the system dynamics subject to an external periodic force and thermal noise and show how the input signal can be amplified by thermal noise through the mechanism of SR. For typical SR to occur, one needs three ingredients (nonlinearity or a threshold, a certain amount of noise, and a subthreshold periodic signal) and a frequency-scale or time-scale matching condition to be satisfied. The results in the preceding sections show that the optomechanical nonlinearity provides a bistable structure for state switching. Thermal noise exists naturally in the system and its amplitude can be tuned by controlling the operating temperature. The only remaining ingredient is a subthreshold signal, which is inadequate for activating bistable state switching. For instance, as seen in Fig.~\ref{SR}(c), the mechanical force with amplitude $F_s=1.5\omega_m$ and frequency $f_s=0.05\omega_m$ causes only a small-amplitude modulation of the self-sustained oscillations in the upper branch without driving it to the lower branch. The SR matching condition is usually that the average period of the noise-induced switching between states should be comparable with half the period of the external signal \cite{Gammaitoni}. The average period of noise-induced random switching is determined by the noise amplitude and the system parameters. Therefore, to satisfy this condition, one can sweep either the noise strength or the system parameters, or sweep the signal. Here, we choose to fix the signal frequency and sweep the thermal noise over a large range and search for an optimal noise strength for matching.

In Fig.~\ref{SR}(a), we plot the signal-to-noise ratio (SNR) [dB] as a function of the noise strength $D_m$. Here, the SNR is defined as the ratio of the spectrum peak to the spectrum background at the central frequency of the peak. We see that the SNR curve peaks at a nonvanishing finite noise strength, which is a characteristic feature of SR. We choose three points (A, B, and C) and study the spectral and temporal properties at different noise levels. The spectra for points A--C are shown in Fig.~\ref{SR}(b), where there are two peaks (a narrow peak at low frequency and a wide peak at high frequency) in every spectrum, unlike the single-peak SNR curve for typical SR. The narrow peak is at a frequency of around $0.008\omega_m$, consistent with the signal frequency $f_s/(2\pi)$. Therefore, this peak is the SR peak due to synchronization to the external signal $F_s$. The wide peak is centered at around $0.15\omega_m$, almost the same as the peak position in Fig.~\ref{switch}, and corresponds to the frequency of the intrinsic self-sustained oscillations in the high-amplitude state.

We turn next to the temporal performance of the system with different noise levels (points A--C). For reference, we  plot in Fig.~\ref{SR}(c) the system dynamics (i.e., the mechanical position $x$) in the absence of noise. The signal clearly involves a slow modulation of the self-sustained oscillations on the upper branch without switching to the lower stable branch, which indicates that the signal is below threshold. With the lowest noise level (point A), the switching between the oscillating and stable states is not particularly regular. When the noise is increased to the intermediate level (point B), we observe excellent switching regularity that is synchronized well with the signal. Increasing the noise further to the highest level (point C), the switching becomes much noisier because of the strong noise background, resulting in the reduction in the SR peak.

\section{Conclusion}
We studied noise-induced CR and SR in an optomechanical system with self-sustained oscillation dynamics. We analyzed the stability properties theoretically and showed the existence of an overlapping region of bistability and self-sustained oscillations. In the vicinity of this region, we found the occurrence of CR that could sustain long-term quasi-coherent oscillations of the system, a result that is the first report of CR in cavity optomechanics. Inside the overlapping region and with the imposition of a subthreshold signal with a matching condition, we observed by numerical means thermal noise-induced periodic switching between self-sustained oscillations and equilibrium, resulting in the external signal being amplified. Unlike the single-peak spectrum of typical SR, there were two resonance peaks in the power spectrum, with the additional peak corresponding to the self-sustained oscillations. Our results shed further light on the interplay between noise and nonlinearity in an optomechanical system and provide a new means for understanding SR.

\section*{Acknowledgment}
The authors thank Dr Zhenglu Duan for helpful discussions.
\section*{Funding}
National Natural Science Foundation of China (No.11504145 and No.11664014); Natural Science Foundation of Jiangxi Province (No.20161BAB211013 and No.20161BAB201023).

\bibliographystyle{unsrt}
\bibliography{Reference}

\end{document}